\documentclass{fundam}

\usepackage[ruled]{algorithm2e}
\usepackage{amsmath}
\usepackage{amssymb}
\usepackage{booktabs}
\usepackage{colortbl}
\usepackage{graphicx}
\usepackage{xspace}
\usepackage{url}
\usepackage{pgfplots}
\pgfplotsset{compat=1.17}

\usepackage{kCol}

\title{Speeding Hirschberg Algorithm for Sequence Alignment}

\author{David Llorens\\
Institute of New Image Technologies\\
Universitat Jaume I\\
SPAIN
\and
Juan Miguel Vilar\corresponding\\
Institute of New Image Technologies\\
Universitat Jaume I\\
SPAIN
}

\runninghead{D. Llorens, J.M. Vilar}{Speeding Hirschberg Algorithm for Sequence Alignment}

\begin{document}
\renewcommand{\O}{{\cal O}}%
\maketitle
\begin{abstract}
  The use of Hirschberg algorithm reduces the spatial cost of recovering the
  Longest Common Subsequence to linear space. The same technique can
  be applied to similar problems like Sequence Alignment. However, the
  price to pay is a duplication of temporal cost. We present here a
  technique to reduce this time overhead to a negligible amount.
\end{abstract}

\section{Introduction}

The answer to many Dynamic Programming problems includes not only the
score of the optimal solution but also the sequence of decisions that
led to it. For example, given two sequences, it is usually more
interesting to find the best alignment between them than the score of
that alignment. This poses a problem since the memory costs associated
to the recovery of the alignment using a direct Dynamic Programming
approach are much higher than the costs of finding the score of that
alignment. Concretely, when aligning two sequences of length~$m$
and~$n$, the score of the alignment can be obtained in $\O(mn)$
steps. Storing the trellis (or at least backpointers) for recovering
the actual alignment implies using $\O(mn)$ memory space.

To remedy this, it is possible to use a technique due to
Hirschberg~\cite{hirschberg75} and popularized for biological
alignments by Myers and Miller~\cite{myers1988} that only
requires~$\O(n+m)$ memory at the cost of duplicating the execution
time. This technique was developed for solving the Longest Common
Subsequence problem and for that particular problem it can be
accelerated using bit vector algorithms since the difference in the
contents of neighbor cells in the trellis is
0~or~1~\cite{Crochemore2003}. This is however not applicable to other
problems like Sequence Alignment.

Here we present a novel approach that reduces the time overhead
to a small fraction of the original with a memory cost that is
still~$\O(n+m)$. Hirschberg's method combines a forward pass of the
first half of the trellis with a backward pass of the second to
find one intermediate point and recurses to find the optimal path.
This reduces memory cost with a time overhead factor of~2.
Our proposal is to use $k$ intermediate points and find them in
a forward pass. This leads to a time overhead factor of
$\frac{k}{k-1}$.

\section{Sequence Alignment}

The problem we will tackle is Sequence alignment. Let there be
given two sequences~$a=\sub{a}{1}{m}$ and~$b=\sub{b}{1}{n}$. Assume
all symbols of those sequences belong to an
alphabet~$\alphabet$. Let~$\g$ be a new symbol not belonging
to~$\alphabet$ that represents a \emph{gap}. An alignment of~$a$
and~$b$ is a pair of sequences~$(a', b')$ such that both $a'$ and $b'$
have symbols from $\alphabet\cup\{\g\}$, both have the same length,
and such that the sequences obtained by taking out the gaps from~$a'$
and~$b'$ are $a$ and $b$, respectively. For instance, given the
sequences of DNA bases, $a=\dna{ACCACTA}$ and~$b=\dna{ACGATC}$, the
pair $(\dna{ACCACTA}, \dna{ACGA\g{}TC})$ is an alignment. It is
possible to score an alignment by defining a similarity measure
between symbols and then adding the similarities of the corresponding
symbols in the two sequences of the alignment. For instance, let us
assign a score of~$+2$ to a pair of equal symbols and~$-1$ to a pair
of different symbols. The previous alignment has then a score of~$5$
as seen below:

\begin{center}
\begin{alignment}{8}
   A &  C &  C &  A &  C &  T &  A\\
   A &  C &  G &  A & \g &  T &  C\\
  +2 & +2 & -1 & +2 & -1 & +2 & -1
\end{alignment}
\end{center}

Obviously, there are many other possible alignments, for example:
\begin{center}
\begin{alignment}{7}
   A &  C &  C &  A &  C &  T &  A\\
   A &  C &  G &  A &  T & \g &  C\\
  +2 & +2 & -1 & +2 & -1 & -1 & -1
\end{alignment}
\end{center}
which has a score of~$2$. We are interested in finding one alignment
with the maximum score (in general, there could be more than one, in
that case, any will do). In our example, the first alignment presented
happens to be optimal.

The usual approach to finding the best alignment is to use Dynamic
Programming, this technique can be found in any book on algorithms
like~\cite{cormen09}. In our case, define $A(i, j)$ to be the score of
the best alignment between the prefixes $\sub{a}{1}{i}$ and
$\sub{b}{1}{j}$ (when~$i$ or~$j$ is zero, we assume that the
corresponding prefixes are empty). It is easy to see
that:

\begin{equation}
  \label{eq:alignment}
  A(i,j) =
  \begin{cases}
     0, & \text{if $i=0$ and $j=0$,} \\
     A(i-1, 0) + s(a_i, \g), &  \text{if $i\neq0$ and $j=0$,} \\
     A(0, j-1) + s(\g, b_j), &  \text{if $i=0$ and $j\neq0$,} \\
     \begin{aligned}
     \max\{&A(i-1,j) + s(a_i, \g), \\
           &A(i, j-1) + s(\g, b_j),\\
           &A(i-1, j-1)+s(a_i, b_j)\},
     \end{aligned}
         & \text{otherwise.}
  \end{cases}
\end{equation}
Where $s$ is the similarity between two symbols\footnote{This equation
does not allow the alignment of two gaps, which makes little sense in
this context. Through the rest of the paper, we will assume that
$s(\g, \g) = -\infty$.}. Then, the score of the best alignment is just
$A(m,n)$. A direct recursive implementation of~$A$ will be
prohibitively costly, but using memoization it can be computed in
$\O(mn)$ time. Just consider $A$ as a matrix (usually called
\emph{trellis}) and fill it in. Also, create a matrix $BP$ of back
pointers that stores for each pair $(i, j)$ the argument that achieved
the maximum in Equation~(\ref{eq:alignment}).

A simple iterative approach suffices:

\begin{algorithm}[h!]
  \caption{Iterative version of the alignment algorithm}
  \KwIn{Two strings: $a$ and $b$ of length $m$ and $n$}
  \KwOut{The optimal alignment of $a$ and $b$}
  $A \gets \Matrix(m, n)$\;
  $BP \gets \Matrix(m, n)$\;
  $A[0, 0]\gets 0$\;
  $BP[0, 0]\gets (-1, -1)$\;
  \For{$i\gets1$ \KwTo $m$}{
     $A[i, 0]\gets A[i-1,0] + s(a_i, \g)$\;
     $BP[i, 0]\gets (i-1, 0)$\;
   }
  \For{$j\gets 1$ \KwTo $n$}{
     $A[0, j]\gets A[0,j-1] + s(\g, b_j)$\;
     \For{ $i\gets1$ \KwTo $m$}{
        $A[i, j]\gets \Maximize\{(i-1, j), (i, j-1), (i-1, j-1)\}$\;
        $BP[i,j]\gets \Choose\{(i-1, j), (i, j-1), (i-1, j-1)\}$\;
      }
    }
  \KwRet $\RecoverAlignment(a, b, BP)$\;
\end{algorithm}

Here, we use \Maximize and \Choose to represent the
maximization process and the election of the predecessor. In this
case, \Maximize is
\begin{equation*}
  \begin{split}
    \Maximize\{(i-1, j), (i, j-1), (i-1, j-1)\} =
         \max\{&A[i-1,j] + s(a_i, \g),\\
               &A[i, j-1] + s(\g, b_j),\\
               &A[i-1, j-1]+s(a_i, b_j)\}
  \end{split}
\end{equation*}
and \Choose returns the pair corresponding to the maximum. Obviously,
they are not implemented as separate functions and the corresponding
values are computed simultaneously in actual code.

The function \RecoverAlignment recovers the alignment from the back
pointers by simply following them until it finds $(-1, -1)$.

The execution of that algorithm in our example stores the following
scores in~$A$:
\begin{center}
\includegraphics[height=5cm]{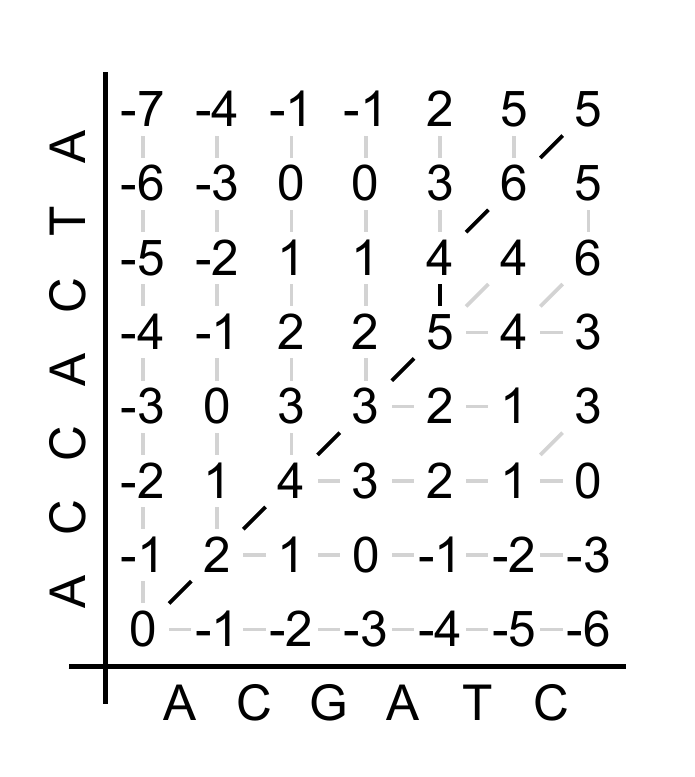}
\end{center}

The lines between the scores of the cells represent the back
pointers. The darker path is the optimum and corresponds to the first
alignment of the example.

All in all, the total spatial cost is $\O(mn)$, which for sequences of
more than a few thousand elements is too high. A first step to reduce
the cost is the observation that the internal loop (that of~$i$) only
uses scores from the previous column, so a simple trick suffices to
eliminate the need to keep~$A$~\cite{cormen09}. Only two columns are
kept ($c$, which is the current column, and $p$, the previous) and
swapped as needed. Algorithm~\ref{alg:reducedSpace} returns the
alignment score using only~$\O(m)$ space.

\begin{algorithm}
  \caption{Reduced space version of the alignment algorithm}
  \label{alg:reducedSpace}
  \KwIn{Two strings: $a$ and $b$ of length $m$ and $n$}
  \KwOut{The optimal alignment of $a$ and $b$}
  \label{alg:column}
  $c \gets \Vector(m)$\;
  $p \gets \Vector(m)$\;
  $c[0]\gets 0$\;
  \For{ $i\gets1$ \KwTo $m$ }{
     $c[i]\gets c[i-1] + s(a_i, \g)$\;
  }
  \For{ $j\gets 1$ \KwTo $n$ }{
     $p, c\gets c, p$\;
     $c[0]\gets p[0] + s(\g, b_j)$\;
     \For{ $i\gets1$ \KwTo $m$ }{
        $c[i]\gets \Maximize\{(i-1, j), (i, j-1), (i-1, j-1)\}$\;
     }
  }
  \KwRet $c[n]$\;
\end{algorithm}
Now \Maximize is slightly changed. Taking into account that
$j-1$ corresponds to~$p$ and~$j$ to~$c$, we obtain:

\begin{equation*}
  \begin{split}
    \Maximize\{(i-1, j), (i, j-1), (i-1, j-1)\} =
          \max\{& c[i-1] + s(a_i, \g),\\
                & p[i] + s(\g, b_j),\\
                & p[i-1]+s(a_i, b_j)\}
  \end{split}
\end{equation*}

Unfortunately, it is not possible to do the same with the
backpointers. The whole $BP$ matrix is needed to recover the
alignment. In the following section we will see how to use the
proposal of Hirschberg to recover the alignment using linear memory at
the cost of duplicating the time.

\section{Hirschberg's method}

Although Hirschberg presented his idea for the problem of finding a
maximal common subsequence of two sequences~\cite{hirschberg75}, it
can be adapted for the case of the alignment.

The idea of the method is to find the point of the alignment path that
passes through the middle column of the trellis. Let~$\pi =
\pi_1\ldots\pi_l\ldots\pi_{|\pi|}$ be an optimal path in the trellis
for the alignment of~$a$ and~$b$. Each~$\pi_l$ is a pair~$(i, j)$ that
marks the alignment of the corresponding prefixes of both sequences.
Now consider the prefix $\sub{\pi}{1}{l}$. It is an optimal path
for the prefixes~$\sub{a}{1}{i}$ and~$\sub{b}{1}{j}$ since otherwise
there would be path~$\rho$ for those prefixes with better score and
then the path~$\rho\sub{\pi}{l+1}{|\pi|}$ would be better than~$\pi$.
A similar reasoning leads to the observation that
$\sub{\pi}{l+1}{|\pi|}$ is an optimal path for $\sub{a}{i+1}{n}$
and~$\sub{b}{j+1}{m}$. Now, let~$\pi_c=(n/2, i)$ be a point of~$\pi$
in column $n/2$ like in Figure~\ref{fig:cutPoint}, left.

\begin{figure}
  \centering
  \includegraphics{manyCuts.1}\qquad\qquad
  \includegraphics{manyCuts.2}
  \caption{The cut point for a trellis (left) and the corresponding
  back pointers (right).}
  \label{fig:cutPoint}
\end{figure}

The key observation is that the total score of the alignment can be
computed in two steps:
\begin{itemize}
\item First, compute the~$c$ column for the alignment of $a$ with the
  first half of $b$, i.e.~$\sub{b}{1}{n/2}$.
\item Second, do the same for~$a$ reversed
  ($a^R=a_ma_{m-1}\ldots{}a_{1}$) and the reversed second half
  of~$b'=b_nb_{n-1}\ldots{}b_{n/2+1}$. This gives a second column, $c'$.
\item Combine $c$ and $c'$ by adding the value of~$c[i]$
  to~$c'[m-i+1]$. The maximum of those values is the score of the
  alignment.
\end{itemize}

In our example, the two alignments are:
\begin{center}
  \hfill%
  \includegraphics[height=5cm]{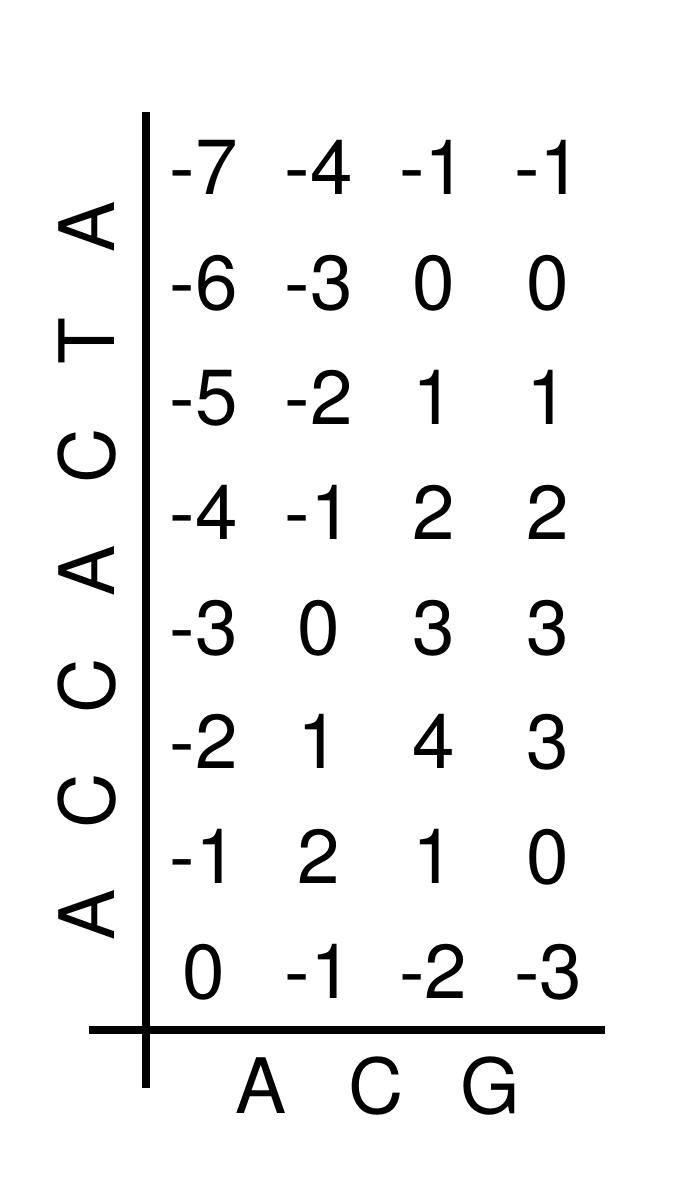}%
  \hfill%
  \includegraphics[height=5cm]{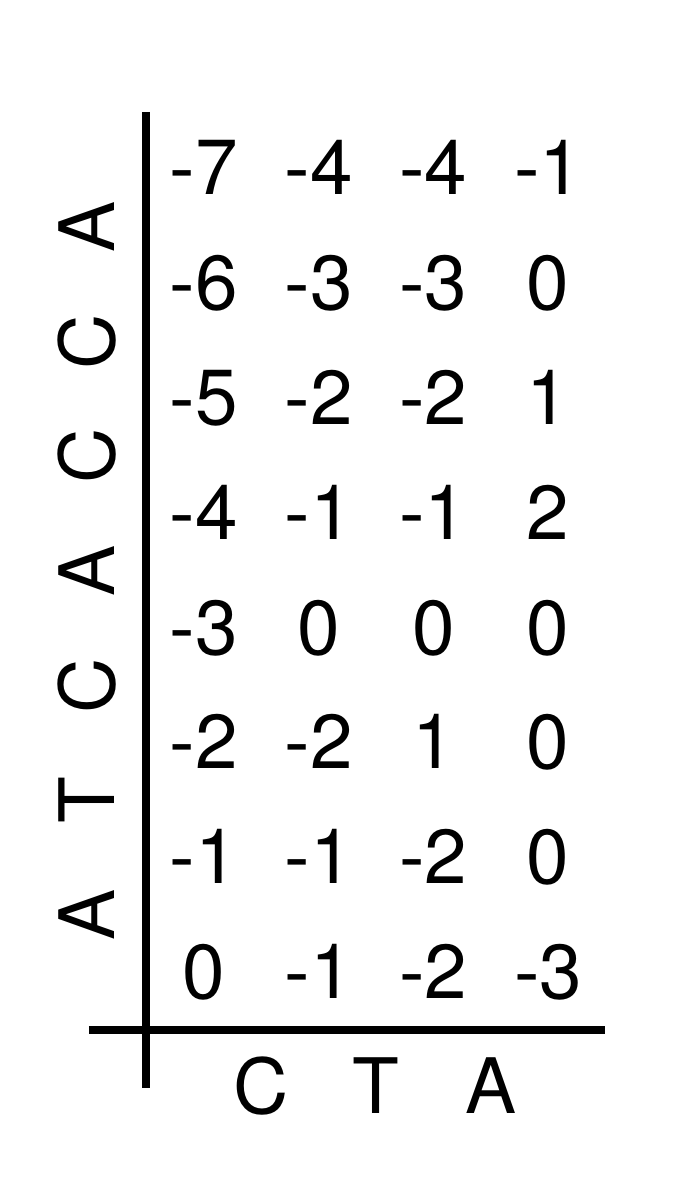}%
  \hfill\rule{0pt}{0pt}%
\end{center}

Therefore, $c$, $c'$ and their combination are:
\begin{center}
  \begin{tabular}{l*{8}{r}}
    \toprule
    $c$             & -3 & 0 & 3 & 3 & 2 & 1 & 0 & -1 \\
    $c'$ (reversed) & -1 & 0 & 1 & 2 & 0 & 0 & 0 & -3 \\
    \midrule
    combination     & -4 & 0 & 4 &\cellcolor{yellow}5 & 2 & 1 & 0 & -4 \\
    \bottomrule
  \end{tabular}
\end{center}

With this, we have found two facts: that the best alignment has a
score of~$5$ and, more importantly, that it aligns \dna{ACC} with
\dna{ACG} and \dna{ACTA} with \dna{ATC}. So, to recover the rest of
the alignment, it suffices to recursively apply the same procedure to
the pairs $(\dna{ACC}, \dna{ACG})$ and $(\dna{ACTA}, \dna{ATC})$. Note
that the spatial cost has been reduced to $O(m + n)$.

We can recover the alignment as shown in
Algorithm~\ref{alg:hirschberg}.

\begin{algorithm}
  \caption{Version of Hirschberg's algorithm for sequence alignment}
  \label{alg:hirschberg}
  \KwIn{Two strings: $a$ and $b$ of length $m$ and $n$}
  \KwOut{The optimal alignment of $a$ and $b$}
  \uIf{$n \leq 1$}{
      \KwRet $\TrivialAlignment(a, b)$\;
      }
  \Else{
      $c\gets \LastColumn (a, \sub{b}{1}{n/2})$\;
      $c'\gets \LastColumn (a^R, b_nb_{n-1}\ldots{}b_{n/2+1})$\;
      $i \gets \argmax_{i} \Combine(c, c')[i]$\;
      \tcp{Recurse on the two halves}
      $(al_1, bl_1)\gets \Alignment(\sub{a}{1}{i}, \sub{b}{1}{n/2})$\;
      $(al_2, bl_2)\gets \Alignment(\sub{a}{i+1}{n}, \sub{b}{n/2+1}{n})$\;
      \KwRet $(al_1al_2, bl_1bl_2)$\;
  }
\end{algorithm}

Here, \LastColumn simply computes the alignment score like in
Algorithm~\ref{alg:column} and returns the last column instead
of only the value in its end. It is interesting to note that
Hirschberg's method disposes of the backpointers completely.

The other function, \TrivialAlignment deals with two simple cases:
when~$b$ is the empty string or has only one symbol. The alignment
corresponding to the first is simply the string~$a$ with $n$ gaps and
for the second case there are two possibilities: either $b_1$ is
aligned to a gap or to one of the symbols of~$a$. This is reflected
in Algorithm~\ref{alg:trivial}.

\begin{algorithm}
  \caption{Algorithm \texttt{TrivialAlignment}}
  \label{alg:trivial}
  \KwIn{Two strings: $a$ and $b$ of length $m$ and $n$ with $n \leq 1$}
  \KwOut{The optimal alignment of $a$ and $b$}
  \uIf{$n = 0$}{
      \KwRet $(a, \rep{\g}{m})$\;
    }
  \Else(\tcp{$n = 1$}){
      \uIf{$\max_{i} \big(s(a_i, b_1) - s(a_i, \g)\big) < s(\g, b_1)$}{
         \KwRet $(\g{}a, b\rep{\g}{m})$\;
      }
      \Else{
         $i\gets\argmax_{i} \big(s(a_i, b_1) - s(a_i, \g)\big)$\;
         \KwRet $(a, \rep{\g}{i-1}b\rep{\g}{n-i})$\;
       }
  }
\end{algorithm}

To find the temporal cost, first note that it is dominated by the cost
of the calls to \LastColumn which is linear in $mn$, so it can be
approximated as $lmn$. Similarly, the call to \TrivialAlignment
has a cost linear in~$m$, and we approximate it as $l'm$. Note also
that the recursive calls only touch the shadowed regions of
Figure~\ref{fig:cutPoint}, and that the total area of these regions is
$\frac{mn}{2}$, no matter the value of~$i$. Therefore, we arrive to
the following recurrence:
\begin{equation}
  \label{eq:hirschbergCost}
  C(n,m) =
  \begin{cases}
    l'm, & \text{if $n=1$,} \\
    lmn + C(n/2, m), & \text{if $n>1$.}\\
  \end{cases}
\end{equation}
This can be easily solved assuming $n$ to be a power of two:
\begin{equation*}
  \begin{split}
    C(n,m) &= lnm + C(n/2, m) = lnm + \frac{lnm}{2} + C(n/2^2, m)\\
    &=\cdots= l'm + \sum_{i=0}^{\log_2n}\frac{lnm}{2^i} \approx 2lnm.
  \end{split}
\end{equation*}

Therefore, the cost of having linear space is the duplication of the
execution time.

\section{The \kcol method}

Now, we will show how can we reduce the overhead of path recovery so
that the total time will be similar to that of using the full trellis
of backpointers but without the memory overhead. Let's return to
Figure~\ref{fig:cutPoint}. Remember that the key issue is to find the
position of~$\pi_c$. We want to do that in a forward pass of the
algorithm.  We need a backpointer from the top right position of the
trellis indicating the row of~$\pi_c$, since we already know the column
($n/2$). With this backpointer it will be possible to perform the
recursive call.  We will also keep a backpointer from $\pi_c$ to the
corner for its recursive call. Although this backpointer will always
be~$0$, we keep it to ease the formulation of the full algorithm. The
diagram in the left of Figure~\ref{fig:cutPoint} represents those
pointers.

Let us consider how can we find $\pi_c$. First, note that this is the
point in which the best path from $(m, n)$ to $(0, 0)$ cuts the column
at $\frac{n}{2}$. Define $bp_j(i)$ as the row in which the best path
from $(i,j)$ to $(0,0)$ cuts column $\frac{n}{2}$. Clearly, the value
of $bp_{\frac{n}{2}}(i)=i$, And for a $j$ greater than $\frac{n}{2}$
the value of $bp_{j+1}(i)$ is equal to the value of $bp_{j'}{i'}$
where $(i', j')$ is the predecessor of $(i, j)$. Therefore a forward
version of Hirschberg's algorithm can be implemented by initializing a
column vector in column $\frac{n}{2}$ and updating it. As mentioned
above, to keep the algorithm uniform, a column will be also
initialized to zeros in the first postion.

With this, we arrive to Algorithm~\ref{alg:1col}. In it, the variable
$bpc$ keeps the backpointers of the current column and $bpp$ those of
the previous column. The function \Maximize has the same form as in
Algorithm~\ref{alg:column}. And assuming $(i',j')$ is the predecessor
of $(i, j)$ in the optimal path, the value of \Choose is:

\begin{equation*}
  \Choose\{(i-1, j), (i, j-1), (i-1, j-1)\} =
  \begin{cases}
    bpc[i-1], & \text{if $(i-1, j)=(i', j')$,}\\
    bpp[i], & \text{if $(i, j-1)=(i', j')$,}\\
    bpp[i-1], & \text{if $(i-1, j-1)=(i', j')$.}\\
  \end{cases}
\end{equation*}

\begin{algorithm}
  \caption{Forward version of Hirschberg algorithm}
  \label{alg:1col}
  \KwIn{Two strings: $a$ and $b$ of length $m$ and $n$}
  \KwOut{The optimal alignment of $a$ and $b$}
  \If{ $n \leq 1$ }{
      \KwRet $\TrivialAlignment(a, b)$\;
    }

  $p \gets\Vector(m)$; $bpp \gets\Vector(m)$\;
  $c \gets\Vector(m)$; $bpc \gets\Vector(m)$\;
  $cols \gets\emptyset$\;
  $c[0]\gets 0$\;
  \For{$i\gets1$ \KwTo $m$}{
     $c[i]\gets c[i-1] + s(s_i, \g)$\;
     $bpc[i]\gets 0$\;
   }
  \For{$j\gets 1$ \KwTo $n$}{
     $c[0]\gets p[0] + s(\g, b_j)$\;
     \For{$i\gets1$ \KwTo $m$}{
        $c[i]\gets \Maximize\{(i-1, j), (i, j-1), (i-1, j-1)\}$\;
        $bpc[i]\gets \Choose\{(i-1, j), (i, j-1), (i-1, j-1)\}$\;
      }
     $p\gets c$\;
     \uIf{ $j = n/2$ \KwOr $j=n$ }{
         $\Push(cols, bpc)$\;
         \lFor{$i\gets1$ \KwTo $m$}{$bpp[i]\gets i$}
       }
    \Else{
         $bpp, bpc\gets bpc, bpp$\;
       }
     }
  $c \gets \Pop(cols)$\;
  $i \gets c[n]$\;
  \tcp{Recurse on the two halves}
  $(al_2, bl_2) \gets \Alignment(\sub{a}{i+1}{n}, \sub{b}{n/2+1}{n})$\;
  $(al_1, bl_1) \gets \Alignment(\sub{a}{1}{i}, \sub{b}{1}{n/2})$\;
  \KwRet $(al_1al_2,bl_1bl_2)$\;
\end{algorithm}

\begin{figure}
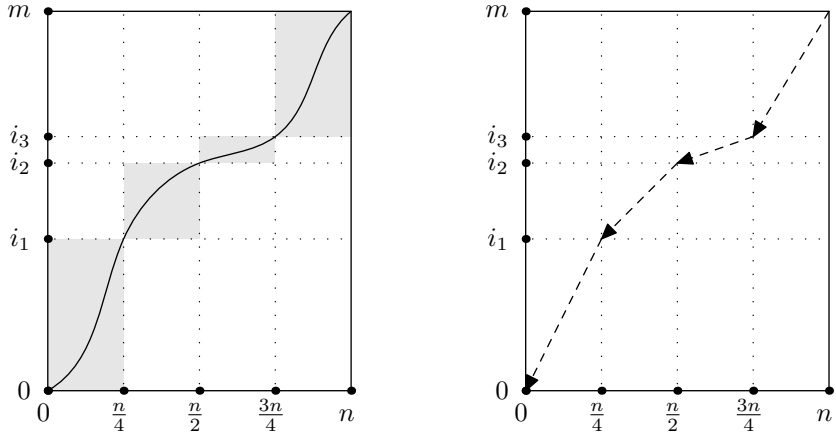

  \centering
  \includegraphics{manyCuts.3}\qquad\qquad
  \includegraphics{manyCuts.4}
  \caption{The cuts with three columns (left) and the corresponding
  back pointers (right).}
  \label{fig:manyCuts}
\end{figure}

Algorithm~\ref{alg:1col} does not give any reduction of
costs over Hirschberg's method but it opens the possibility of storing
more intermediate columns, like in Figure~\ref{fig:manyCuts}.

This gives rise to Algorithm~\ref{alg:kcol}.

\begin{algorithm}{}
  \caption{The algorithm \kcol}
  \label{alg:kcol}
  \KwIn{Two strings: $a$ and $b$ of length $m$ and $n$}
  \KwOut{The optimal alignment of $a$ and $b$}
  \If{$n \leq 1$}{
      \KwRet $\TrivialAlignment(a, b)$\;
  }
  $p \gets\Vector(m)$; $bpp \gets\Vector(m)$\;
  $c \gets\Vector(m)$; $bpc \gets\Vector(m)$\;
  $cols \gets\emptyset$\;
  $c[0]\gets 0$\;
  \For{$i\gets1$ \KwTo $m$}{
     $c[i]\gets c[i-1] + s(s_i, \g)$\;
     $bpc[i]\gets 0$\;
  }
  \For{$j\gets 1$ \KwTo $n$}{
     $c[0]\gets p[0] + s(\g, b_j)$\;
     \For{ $i\gets1$ \KwTo $m$}{
        $c[i]\gets \Maximize\{(i-1, j), (i, j-1), (i-1, j-1)\}$\;
        $bpc[i]\gets \Choose\{(i-1, j), (i, j-1), (i-1, j-1)\}$\;
     }
     $p\gets c$\;
     \If{$\IsSpecial(j)$}{
         $\Push(cols, bpc)$\;
         \lFor{$i\gets1$ \KwTo $m$}{$bpp[i]\gets i$}
       }\Else{
         $bpp, bpc\gets bpc, bpp$\;
     }
  }
  $i\gets n$\;
  \For{$j\gets k$ \KwDownTo $1$}{
      $c\gets \Pop(cols)$\;
      $i'\gets c[i]$\;
      $(al_j, bl_j) \gets \Alignment(\sub{a}{i'+1}{i}, \sub{b}{(j-1)*n/k+1}{j*n/k})$\;
      $i\gets i'$\;
  }
  \KwRet $(\sub{al}{1}{k},\sub{bl}{1}{k})$;
\end{algorithm}
The function $\IsSpecial$ returns true when the column is a
special one, i.e. is of the form $\frac{ln}{k}$ adequately
rounded. When that is the case, the current backpointer column is
pushed in $cols$.

To understand the reduction on the cost, note that the total area of
the shadowed regions in Figure~\ref{fig:manyCuts}, left, is
$\frac{nm}{k}$, which can be substantially lower than Hirschberg's
$\frac{nm}{2}$ for small values of~$k$. To compute the cost of
Algorithm~\ref{alg:kcol}, we can use the following recursion:

\begin{equation}
  \label{eq:costK}
  C(n, m) =
  \begin{cases}
    t'm,              &\text{if $n=1,$}\\
    tnm + C(n/k, m), &\text{if $n> 1.$}
  \end{cases}
\end{equation}
If we expand the recursion we arrive to:
\begin{equation*}
  \begin{split}
    C(n,m) &= tnm + C(n/k, m) = tnm + \frac{tnm}{k} + C(n/k^2, m)\\
    &=\cdots= t'm \sum_{i=0}^{\log_kn}\frac{nm}{k^i} \approx \frac{k}{k-1}tnm.
  \end{split}
\end{equation*}

So the total cost depends on a constant~$t$ that will be slightly
higher than the~$l$ of Hirschberg for the need of managing the
backpointers and the fraction $\frac{k}{k-1}$ that rapidly
approaches~$1$. So as long as the ratio of $t$ to $l$ is kept lower
than two, the overhead will be very low with a memory cost of a
modest~$\O(km)$.

\section{Experiments}
\label{sec:experiments}

To test \kcol, we aligned two sequences corresponding to the human and
mouse versions of the Titin protein. Each sequence has approximately
35 thousand aminoacids. The matrix used to evaluate the distance
between the aminoacids was \emph{blosum 62}.

\paragraph{Implementation Considerations}

The different algorithms were coded in \emph{rust} and their
implementation is freely available in github
(\url{https://github.com/DavidLlorens/SequenceAlignment}). The election
of the programming language was motivated by efficiency considerations
and to avoid the influence of garbage collectors.

All the executions were done in a Linux workstation equipped with an
Intel Core i7-7700 CPU and 32Gb of RAM. The times were measured by the
program using the \texttt{Instant} structure of rust library
\texttt{std::time}. Memory occupation was measured as the maximum
resident size reported by the \texttt{time} utility of GNU. All the
results presented are the averages of ten executions of the program.

Some improvements were done over the naive implementation of the
algorithms as presented. 

The recursive algorithms used the basic quadratic memory
implementation when the number of cells in the trellis was small
enough. Different number of nodes were considered as recursion base
size and their impact on the running time and memory usage was
analyzed.

Another optimization was the use of a single column of the trellis
together with a few auxiliary variables. Since only two cells from
the previous column are needed to compute the value of a cell in the
current column, a bit of ``juggling'' with the variables allows the
saving of the memory associated to a column.

A final optimization was also used in \kcol to reduce cache misses.
The vector corresponding to the column of the trellis and the vector
with the backpointers where joined in a single vector of pairs.

\paragraph{Baseline Results}

Two baseline measures were considered. First, the cost of finding
only the score without the alignment. Second, the use of the quadratic
space algorithm (i.e. storing the whole trellis) to recover the
alignment.

The results can be seen in this table:\bigskip

\begin{tabular}{lrr}
  \toprule
  Algorithm & time (s) & size (KB) \\
  \midrule
  Only score & 1.55 & 2,736 \\
  Quadratic space & 2.66 & 1,185,988\\
  \bottomrule
\end{tabular}\bigskip

It is clear that recovering the alignment has a cost in time and
memory.

\paragraph{Effect of the Recursion Base Size}

Analyzing Figure~\ref{fig:hirschbergResource} it is clear that using
the quadratic space algorithm when the number of nodes is small enough
saves space and has a negligible effect on time in the case of
Hirschberg's algorithm. Also, it is interesting to see that the
overhead of recovering the alignment is almost the same as the
overhead of the quadratic space algorithm. This can be attributed to
the large number of cache misses incurred in by the quadratic
algorithm while filling the trellis.

The effect of the recursion base size is similar in the case of \kcol.
For example, it can be seen in Figure~\ref{fig:kcol32Resource}, that
for the case $k=32$, once the recursion base size is big enough there
is a reduction in memory used while the effect on the time is
negligible.

\begin{figure}
  \begin{tabular}[c]{c}
    \begin{tikzpicture}
  \pgfplotsset{set layers}
  \begin{semilogxaxis}[
      width = 0.38\linewidth,
      xminorticks = false,
      scale only axis,
      scaled ticks = false,
      axis y line*=left,
      xlabel = Recursion base,
      ylabel = seconds,
      ymin = 1.5,
      ymax = 3.5,
      legend style={at={(0,1.01)}, anchor=south west},
    ]
  \legend{Time}
  \addplot [ mark = x, blue ]
    table [
    x = base,
    y = time,
    ]{
  base time size
  1 2.71 13696
  100 2.68 8330
  300 2.69 7781
  1000 2.68 7308
  3000 2.66 7051
  10000 2.67 6623
  30000 2.67 6442
  100000 2.65 6633
  300000 2.66 6592
  1000000 2.66 6618
  };
  \end{semilogxaxis}
  \begin{semilogxaxis}[
      width = 0.38\linewidth,
      scale only axis,
      scaled ticks = false,
      ylabel = KB,
      axis x line=none,
      axis y line*=right,
      legend style={at={(1,1.01)}, anchor=south east},
    ]
  \legend{Memory}
  \addplot [ mark = *, red, dashed ]
    table [
    x = base,
    y = size,
    ]{
  base time size
  1 2.71 13696
  100 2.68 8330
  300 2.69 7781
  1000 2.68 7308
  3000 2.66 7051
  10000 2.67 6623
  30000 2.67 6442
  100000 2.65 6633
  300000 2.66 6592
  1000000 2.66 6618
  };
  \end{semilogxaxis}
\end{tikzpicture}
  \end{tabular}
\quad
\begin{tabular}[c]{rrc}
\toprule
  R. base & Time & Memory\\
\midrule
          0 & 2.71 & \makebox[0pt][r]{1}3,696\\
        100 & 2.68 &                    8,330\\
        300 & 2.69 &                    7,781\\
      1,000 & 2.68 &                    7,308\\
      3,000 & 2.66 &                    7,051\\
     10,000 & 2.67 &                    6,623\\
     30,000 & 2.67 &                    6,442\\
    100,000 & 2.65 &                    6,633\\
    300,000 & 2.66 &                    6,592\\
  1,000,000 & 2.66 &                    6,618\\
\bottomrule
\end{tabular}

  \caption{Resource usage of the implementation of Hirschberg's
  algorithm as function of the recursion base size.}
  \label{fig:hirschbergResource}
\end{figure}
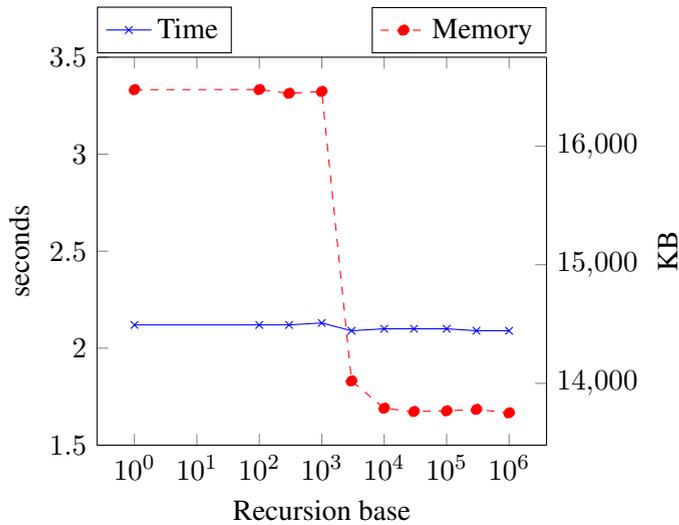

\begin{figure}
  \begin{tabular}[c]{c}
    \begin{tikzpicture}
  \pgfplotsset{set layers}
  \begin{semilogxaxis}[
      width = 0.38\linewidth,
      xminorticks = false,
      scale only axis,
      scaled ticks = false,
      axis y line*=left,
      xlabel = Recursion base,
      ylabel = seconds,
      ymin = 1.5,
      ymax = 3.5,
      legend style={at={(0,1.01)}, anchor=south west},
    ]
  \legend{Time}
  \addplot [ mark = x, blue ]
    table [
    x = base,
    y = time,
    ]{
  k base time size
  32 1 2.12 16477
  32 100 2.12 16478
  32 300 2.12 16446
  32 1000 2.13 16462
  32 3000 2.09 14020
  32 10000 2.10 13790
  32 30000 2.10 13762
  32 100000 2.10 13768
  32 300000 2.09 13780
  32 1000000 2.09 13751
  };

  \end{semilogxaxis}
  \begin{semilogxaxis}[
      width = 0.38\linewidth,
      scale only axis,
      scaled ticks = false,
      ylabel = KB,
      axis x line=none,
      axis y line*=right,
      legend style={at={(1,1.01)}, anchor=south east},
    ]
  \legend{Memory}
  \addplot [ mark = *, red, dashed ]
    table [
    x = base,
    y = size,
    ]{
  k base time size
  32 1 2.12 16477
  32 100 2.12 16478
  32 300 2.12 16446
  32 1000 2.13 16462
  32 3000 2.09 14020
  32 10000 2.10 13790
  32 30000 2.10 13762
  32 100000 2.10 13768
  32 300000 2.09 13780
  32 1000000 2.09 13751
  };
  \end{semilogxaxis}
\end{tikzpicture}
  \end{tabular}
\quad
 \begin{tabular}{rrc}
\toprule
   R. base & Time & Memory\\
\midrule
         0 & 2.12 & 16,477\\
       100 & 2.12 & 16,478\\
       300 & 2.12 & 16,446\\
     1,000 & 2.13 & 16,462\\
     3,000 & 2.09 & 14,020\\
    10,000 & 2.10 & 13,790\\
    30,000 & 2.10 & 13,762\\
   100,000 & 2.10 & 13,768\\
   300,000 & 2.09 & 13,780\\
 1,000,000 & 2.09 & 13,751\\
\bottomrule
\end{tabular}
  \caption{Resource usage of the implementation of \kcol with $k=32$
  as function of the recursion base size.}
  \label{fig:kcol32Resource}
\end{figure}

\paragraph{Effect of the value of $k$} You can see in
Figure~\ref{fig:kcolResource} the results for a recursion base size
of~30,000 nodes. It is clear from that figure that using a value
of~$k$ equal to~16, the increment of time needed to recover the
alignment is reduced from the~1.12s of Hirschberg to~0.61s with a
small increase of memory. Using a value of~$k$ equal to~32, the
increase of memory is still acceptable and the time overhead of
recovering the alignment is reduced to~0.55s.

\pgfplotsset{major grid style={dashed}}

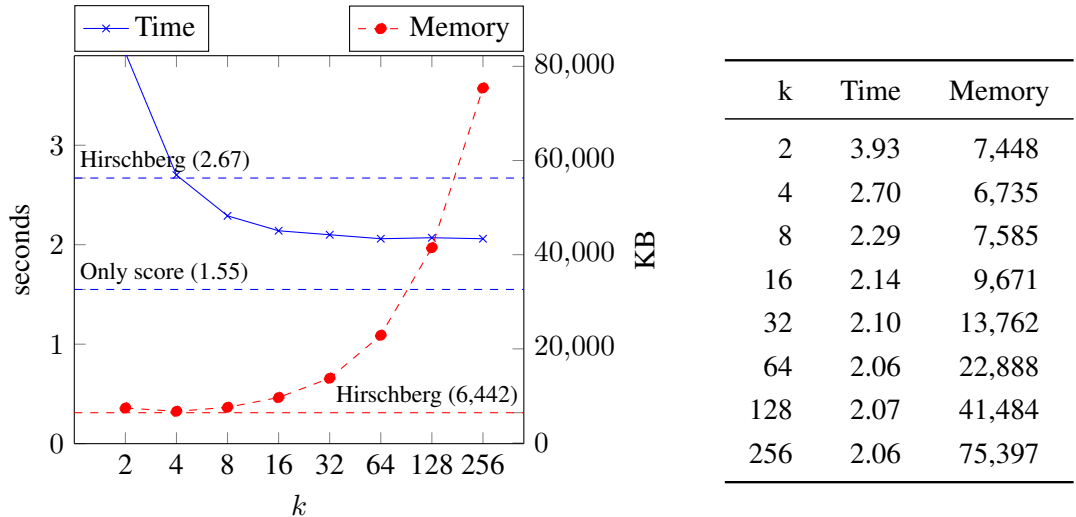
\begin{figure}
  \begin{tabular}[c]{c}
    \begin{tikzpicture}
  \pgfplotsset{set layers}
  \begin{semilogxaxis}[
      width = 0.38\linewidth,
      xminorticks = false,
      xtick = data,
      xticklabels = {2, 4, 8, 16, 32, 64, 128, 256},
      extra y ticks = {1.55,2.67},
      extra y tick style = {grid = none,
                            tick label style =
                             {anchor= south west,
                              outer sep = -1.5pt,
                              font=\footnotesize}},
      extra y tick labels = {Only score (1.55),Hirschberg (2.67)},
      log basis x = 2,
      scale only axis,
      scaled ticks = false,
      axis y line*=left,
      xlabel = {$k$},
      ylabel = seconds,
      xmin = 1,
      ymin = 0,
      ymax = 3.9,
      legend style={at={(0,1.01)}, anchor=south west},
    ]
  \legend{Time};
  \addplot [ mark = x, blue ]
    table [
    x = k,
    y = time,
    ]{
  k base time size
  2 30000 3.93 7448
  4 30000 2.70 6735
  8 30000 2.29 7585
  16 30000 2.14 9671
  32 30000 2.10 13762
  64 30000 2.06 22888
  128 30000 2.07 41484
  256 30000 2.06 75397
  };
  \addplot [blue, update limits = false, dashed]
    coordinates { (1, 2.67) (1024, 2.67) };
  \addplot [blue, update limits = false, dashed]
    coordinates { (1, 1.55) (1024, 1.55) };

  \end{semilogxaxis}
  \begin{semilogxaxis}[
      width = 0.38\linewidth,
      scale only axis,
      scaled ticks = false,
      ylabel = KB,
      axis x line=none,
      axis y line*=right,
      xmin = 1,
      legend style={at={(1,1.01)}, anchor=south east},
      extra y ticks = {6442},
      extra y tick style = {grid = none,
                            tick label style =
                             {anchor= south east,
                              outer sep = -1.5pt,
                              font=\footnotesize}},
      extra y tick labels = {Hirschberg (6{,}442)},
    ]
  \legend{Memory}
  \addplot [ mark = *, red, dashed ]
    table [
    x = k,
    y = size,
    ]{
  k base time size
  2 30000 3.93 7448
  4 30000 2.70 6735
  8 30000 2.29 7585
  16 30000 2.14 9671
  32 30000 2.10 13762
  64 30000 2.06 22888
  128 30000 2.07 41484
  256 30000 2.06 75397
  };
  \addplot [red, update limits = false, dashed]
    coordinates { (1, 6442) (1024, 6442) };
  \end{semilogxaxis}
\end{tikzpicture}
  \end{tabular}
\quad
\def\zz{\phantom{0}}%
\begin{tabular}{rrc}
\toprule
  k   & Time & Memory\\
\midrule
    2 & 3.93 & \zz7,448\\
    4 & 2.70 & \zz6,735\\
    8 & 2.29 & \zz7,585\\
   16 & 2.14 & \zz9,671\\
   32 & 2.10 &   13,762\\
   64 & 2.06 &   22,888\\
  128 & 2.07 &   41,484\\
  256 & 2.06 &   75,397\\
\bottomrule
\end{tabular}
  \caption{Resource usage of the implementation of \kcol
  as function of~$k$ for a recursion base size of 30,000 elements.}
  \label{fig:kcolResource}
\end{figure}

\section{Conclusions}

Recovering the alignment or in general the optimal solution to a
Dynamic Programming problem can involve large spatial costs if a
backpointer structure needs to be kept in memory. Hirschberg's approach
reduces this cost to a linear factor of the input size. This is
achieved by considering where the optimal solution crosses the central
column of the trellis. We have shown
how to further reduce the cost using \kcol, an approach that
keeps the intersections of the optimal solution with $k$ columns of
the trellis. And these intersections are computed during a forward
traversal, instead of the combination of forward and backward
traversals performed by Hirschberg's algorithm.

The experiments presented show that for the problem of protein
alignment the overhead incurred for recovering the optimal solution is
reduced to nearly a half.

\bibliographystyle{fundam}
\bibliography{kCol}

\end{document}